\documentclass{article}

\usepackage{amssymb}
\usepackage{amsthm}
\usepackage{amsmath}
\usepackage{physics}
\usepackage[hyphens]{url}
\usepackage{rotating}
\usepackage{biblatex}
\usepackage{authblk}
\usepackage[margin=1in]{geometry}
\usepackage{url}
\usepackage{breakurl}
\usepackage[breaklinks]{hyperref}

\newtheorem{theorem}{Theorem}[section]
\newtheorem{definition}{Definition}[section]
\newtheorem{lemma}[theorem]{Lemma}

\title{Oracle Separations Between Quantum and Non-interactive Zero-Knowledge Classes}

\author[1]{Benjamin Morrison}
\author[2]{Adam Groce}
\affil[1]{University of New Mexico}
\affil[2]{Reed College}
\bibliography{simple}
\date{}

\begin{document}

\maketitle

\begin{abstract}
We study the relationship between problems solvable by quantum algorithms in polynomial time and those for which zero-knowledge proofs exist.  In prior work, Aaronson \cite{aaronson} showed an oracle separation between BQP and SZK, i.e.~an oracle $A$ such that \(\mathrm{SZK}^A \not\subseteq \mathrm{BQP}^A\).  In this paper we give a simple extension of Aaronson's result to \textit{non-interactive} zero-knowledge proofs with perfect security.  This class, NIPZK, is the most restrictive zero-knowledge class.  We show that even for this class we can construct an $A$ with \(\mathrm{NIPZK}^A \not\subseteq \mathrm{BQP}^A\).
\end{abstract}

\section{Introduction}

We investigate the relationship between quantum-computable problems and those with zero-knowledge proofs.  We are motivated by the general desire of complexity theory to understand all relationships between complexity classes,  as well as implications this particular relationship has for quantum-resilient cryptography.  Specifically we consider the class BQP, those languages decidable by a quantum computer in polynomial time with bounded error.  (See \cite{mermin2007quantum} for a more thorough discussion.)  

Zero-knowledge is really a family of complexity classes.  In a zero-knowledge proof for language $L$, the prover $P$ must convince a verifier $V$ that $x \in L$.  The zero-knowledge property requires that the verifier cannot learn anything other than the statement being proved.  (For example, $P$ cannot send a witness for~$x$.)  This is formalized by requiring that a simulator without access to $P$ can produce a transcript $T'$ that is indistinguishable from a transcript $T$ of a real interaction.  If $T'$ is required to be distributed identically to $T$, then the resulting complexity class is \textit{perfect} zero-knowledge (PZK).  If it is required only to be statistically close, we get \textit{statistical} zero-knowledge (SZK).  If it is only required to be computationally indistinguishable, we get \textit{computational} zero-knowedge (CZK).  (See \cite{goldreichbook} for a more thorough discussion.)

We can further restrict the three classes above by requiring that the protocols be \textit{non-interactive}.  That is, we require that the whole interaction between $P$ and $V$ consist of a single message sent from $P$ to $V$.  To make this possible, we must give the parties access to a common random string.  We therefore have three non-interactive classes, analogous to those above (NICZK, NISZK, and NIPZK).

There are no unconditional results proving anything about the relationship between BQP and any of the six zero-knowledge classes discussed above.  However, Aaronson \cite{aaronson} gave an oracle separation, an oracle $A$ under which  \(\mathrm{SZK}^A \not\subseteq \mathrm{BQP}^A\).  This is evidence that there are problems with zero-knowledge proofs but no quantum algorithms, and it rules out many proof techniques for proving otherwise.  In this paper we give a simple extension of this result, showing an oracle separation between BQP and NIPZK.  NIPZK is the most restrictive of the zero-knowledge classes, so when we show an $A$ such that \(\mathrm{NIPZK}^A \not\subseteq \mathrm{BQP}^A\) we implicitly show the same for PZK, NICZK, and NISZK.  

%However, for the sake of clarity we first present proofs of weaker results.  First we briefly present Aaronson's proof for SZK, making explicit some details that do not appear in his original work.  We do this because we use those details and a similar proof technique for our results.  We then present a proof for NISZK.  This is a new result, but follows almost immediately from other known results.  Finally, we present our proof for NIPZK.

\paragraph{Implications for cryptography}   In recent years a variety of cryptographic protocols have been built using non-interactive zero-knowledge proofs.  For example, Miller \textit{et al.}~use them to create a cryptocurrency that can be generated and spent anonymously \cite{miller-kosba-katz-shi}.  Haralambiev uses them to create leakage-resilient signatures, signatures that remain secure even when some of the secret key is disclosed \cite{haralambiev}.  Juels \textit{et al.}~show they can be used in a less desirable way, allowing crowdfunding to be used to reward hackers for disclosing the secret information of their victims \cite{juels-kosba-shi}.  

The cryptographic community has also recently spent considerable effort finding protocols that will remain secure in the face of adversaries with quantum computers.  If the non-interactive zero-knowledge classes were contained in BQP, it would imply that cryptographic protocols using such proofs could not be made resilient to such quantum adversaries.

\section{Oracle Separation Between SZK and BQP}
Aaronson \cite{aaronson} proved the following result:
\begin{theorem}
\label{separation-aaronson}
There exists an oracle A such that \(\mathrm{SZK}^A \not\subseteq \mathrm{BQP}^A\).
\end{theorem}

Our own proofs follow a similar structure and rely on some of Aaronson's lemmas, so we begin by recalling a few key details of his proof. He begins with a lower bound for the quantum query complexity of the Collision Problem.  The problem is defined as follows.
\begin{definition}[Collision Problem]
Let \(n\) be an integer and \(X~:~\{1,...,n\}~\rightarrow~\{1,...,n\}\), represented in the standard way as a list of outputs. Suppose either \(X\) is one-to-one (that is, each element of \(\{1,...,n\}\) is output for exactly one input) or \(X\) is r-to-one\footnote{It is sufficient for our result to restrict the problem to the $r=2$ case.} for a fixed $r \geq 2$ (that is, each element of \(\{1,...,n\}\) is output for exactly r inputs or not at all.) Then given the ability to query \(X\), the Collision Problem \(\mathrm{Col}^r_n\) is to accept if \(X\) is one-to-one and reject if \(X\) is $r$-to-one.
\end{definition}

Aaronson then shows the following result, which we present without proof.  $Q_2(\cdot)$ represents the (bounded error) quantum \textit{query} complexity of the problem, defined as the number of bits of the input that the algorithm must examine.

\begin{lemma}
\label{collision-aaronson}
\(Q_2\left(\mathrm{Col}_n^2\right) = \Omega\left(n^{1/5}\right)\).
\end{lemma}

Kutin \cite{kutin} proves a stronger version of the collision lower bound, \(\Omega\left((n/r)^{1/3}\right)\), that also applies when $r \neq 2$.  (This result is also a strengthening of the result of Shi \cite{shi2002quantum}, which gives the same bound but requires a larger output set for the function.)  From either result, a diagonalization can be performed to produce the desired oracle separation.

\section{Oracle Separation Between NIPZK and BQP}

We now prove the following new result:

\begin{theorem}
There exists an oracle \(A\) such that \(\mathrm{NIPZK}^A \not\subseteq \mathrm{BQP}^A\).
\end{theorem}

It suffices to demonstrate a NIPZK algorithm for \(\mathrm{Col}^2_n\). The algorithm, inspired by the algorithm for uniformity testing given by Malka \cite{malka}, proceeds as follows. The prover divides the shared random string into two strings \(r_1\) and \(r_2\), each of length \(n\). For each \(r_i\), it chooses uniformly a string \(x_i\) with \(X(x_i) = r_i\). It then sends the chosen \(x_i\) to the verifier. The verifier accepts if \(X(x_i) = r_i\) for both \(i\). 

We now prove the algorithm is NIPZK. First, we prove its completeness. If \(X\) is one-to-one, then its image equals its codomain, and so the \(x_i\) can always be selected to be valid, regardless of the \(r_i\). Thus the verifier will always accept any one-to-one function.

Next, we prove its soundness. If \(X\) is two-to-one, then half of its codomain is not in its image. Thus, with probability \(\frac{3}{4}\), at least one of \(r_1\) or \(r_2\) is not in the image of \(X\). Thus, with probability \(\frac{3}{4}\), the prover cannot select \(x_i\) that the verifier will accept. Thus the soundness error is \(\frac{1}{4}\).

Next, we prove its perfect zero-knowledge property. The simulator can simply randomly pick two inputs \(x_1\) and \(x_2\), then run them through \(X\) to get appropriate \(r_i\).  Since the \(x_i\) are selected uniformly, when $X$ is one-to-one the \(r_i\) are also uniformly distributed. Furthermore, for those \(r_i\), there is only one possible pair of \(x_i\); thus the simulator can exactly recreate the distribution over inputs to the verifier. From there, it can simply perfectly simulate any verifier on those inputs.

Thus \(\mathrm{Col}^2_n \in\) NIPZK, and the theorem follows as above.

\section{Conclusion}
We constructed a NIPZK query algorithm for the collision problem. Using this algorithm and the quantum query lower bound on the collision problem we have demonstrated the existence of an oracle relative to which NIPZK~\(\not\subseteq\)~BQP. This result has applications to the quantum-resistance of cryptography and cryptocurrency, where algorithms occasionally rely on non-interactive zero-knowledge proof protocols. Our result suggests that the use of those proofs does not introduce vulnerabilities into those algorithms in the presence of a quantum adversary.
The next step would be to extend this oracle separation into an algebraic oracle separation~\cite{aaronson-wigderson}, which would rule out a wider array of proof techniques and give additional evidence that~NIPZK~\(\not\subseteq\)~BQP.

\printbibliography

\end{document}